\documentclass[letter]{aa}

\usepackage[breaklinks=true,colorlinks=false,
     linkcolor=blue,
     filecolor=blue,
     citecolor = blue,      
     urlcolor=blue,]{hyperref}
\usepackage{natbib,twoopt}
\usepackage{graphicx}
\usepackage{txfonts}
\bibpunct{(}{)}{;}{a}{}{,}

\begin{document} 

\title{ClG 0217+70: A massive merging galaxy cluster with a large radio halo and relics}

\subtitle{}

\author{
X. Zhang \inst{1,2}
\and A. Simionescu \inst{2,1,3}
\and J. S. Kaastra \inst{2,1}
\and H. Akamatsu \inst{2}
\and D. N. Hoang \inst{4}
\and \\C. Stuardi \inst{5,6}
\and R. J. van Weeren \inst{1}
\and L. Rudnick \inst{7}
\and R. P. Kraft \inst{8}
\and S. Brown \inst{9}
}

\institute{
Leiden Observatory, Leiden University, PO Box 9513, 2300 RA Leiden, The Netherlands \\
\email{xyzhang@strw.leidenuniv.nl}
\and SRON Netherlands Institute for Space Research, Sorbonnelaan 2, 3584 CA Utrecht, The Netherlands
\and Kavli Institute for the Physics and Mathematics of the Universe (WPI), The University of Tokyo, Kashiwa, Chiba 277-8583, Japan
\and Hamburger Sternwarte, University of Hamburg, Gojenbergsweg 112, 21029 Hamburg, Germany
\and Dipartimento di Fisica e Astronomia, Universit\`{a} di Bologna, via Gobetti 93/2, 40122 Bologna, Italy
\and INAF - Istituto di Radioastronomia di Bologna, Via Gobetti 101, 40129 Bologna, Italy
\and Minnesota Institute for Astrophysics, University of Minnesota, 116 Church St. S.E., Minneapolis, MN 55455, USA
\and Harvard-Smithsonian Center for Astrophysics, 60 Garden Street, Cambridge, MA 02138, USA
\and Department of Physics and Astronomy, University of Iowa, 203 Van Allen Hall, Iowa City, IA 52242, USA
}

\abstract
{
We present an analysis of archival \emph{Chandra} data of the merging galaxy cluster ClG 0217+70. The \ion{Fe}{XXV} He$\alpha$ X-ray emission line is clearly visible in the 25 ks observation, allowing a precise determination of the redshift of the cluster as $z=0.180\pm0.006$. We measure $kT_{500}=8.3\pm0.4$ keV and estimate $M_{500}=(1.06\pm0.11)\times10^{15}\ M_\sun$ based on existing scaling relations. Correcting both the radio and X-ray luminosities with the revised redshift reported here, which is much larger than previously inferred  based on sparse optical data, this object is no longer an X-ray underluminous outlier in the $L_\mathrm{X}-P_\mathrm{radio}$ scaling relation. The new redshift also means that, in terms of physical scale, ClG 0217+70 hosts one of the largest radio halos and one of the largest radio relics known to date. Most of the relic candidates lie in projection beyond $r_{200}$. The X-ray morphological parameters suggest that the intracluster medium is still dynamically disturbed. Two X-ray surface brightness discontinuities are confirmed in the northern and southern parts of the cluster, with density jumps of $1.40\pm0.16$ and $3.0\pm0.6$, respectively.
 We also find a $700\times200$ kpc X-ray faint channel in the western part of the cluster, which may correspond to compressed heated gas or increased non-thermal pressure due to turbulence or magnetic fields.}


\titlerunning{Galaxy cluster merger: ClG 0217+70}

\maketitle

\section{Introduction}\label{sec:introduction}
Galaxy cluster mergers are the most extreme events in the universe and can release energies up to $10^{64}$ erg. The shocks and magnetohydrodynamic (MHD) turbulence generated during these mergers heat the intracluster medium (ICM) and can also (re)accelerate particles into the relativistic regime (see \citealt{2014IJMPD..2330007B} for a theoretical review). Synchrotron radiation emitted by these relativistic particles as they gyrate around intergalactic magnetic field lines leads to observed giant radio halos and radio relics (see \citealt{2019SSRv..215...16V} for a review). Merging galaxy clusters therefore provide unique laboratories to study particle acceleration in a high thermal-to-magnetic pressure ratio plasma. Among the large merging galaxy cluster sample, clusters that host double relics are a rare subclass. They usually have a simple merging geometry with the merger axis close to the plane of the sky and therefore suffer less from projection uncertainties. 

\object{ClG 0217+70}, also known as \object{8C 0212+703} \citep{1995MNRAS.274..447H} or \object{1RXS J021649.0+703552}, is a radio-selected merging cluster \citep{1997A&AS..124..259R,2006AN....327..549R}, which hosts several peripheral radio relic candidates located on opposite sides of a central radio halo \citep{2011ApJ...727L..25B}. 
Among the relic candidates, sources C, E, F, and G (see Fig. \ref{fig:image} for definition) are not associated with any optical galaxy, and a recent study shows that their spectral indices are steeper towards the cluster center (Hoang et al. in prep.), indicating a shock acceleration feature.
This cluster appears as an X-ray underluminous outlier in the $L_\mathrm{X}-P_\mathrm{radio}$ scaling relation  \citep[e.g.,][]{2009A&A...507..661B,2013ApJ...777..141C}. 
One possible explanation for this is the misestimation of the redshift; because of the  lack of  deep optical data, this was believed to be $z=0.0655$ \citep{2011ApJ...727L..25B}.
An accurate redshift is essential to scale the physical properties of the cluster and those of the diffuse radio sources (e.g., size and luminosity). 

Here we present an analysis of archival \emph{Chandra} data, which allows a precise determination of the cluster redshift via the ICM Fe-K line. The high spatial resolution of \emph{Chandra} also enables us to search for surface brightness discontinuities related to the merger.
We adopt a $\Lambda$CDM cosmology where $H_0=70$ km s$^{-1}$ Mpc$^{-1}$, $\Omega_\mathrm{m}=0.3$, and $\Omega_\Lambda=0.7$.

\section{Observations and data reduction}\label{sec:obs}
We  analyzed the 24.75 ks \emph{Chandra} archival data (ObsID: 16293). 
The Chandra Interactive Analysis of Observations (CIAO)\footnote{\url{https://cxc.harvard.edu/ciao/}} v4.12 with CALDB 4.9.0 is used for data reduction. The level 2 event file is generated by the task \texttt{chandra\_repro} with the VFAINT mode background event filtering. We extracted the 100 s binned 9 -- 12 keV light curve for the whole field and did not find flares. Therefore, we used the entire exposure period for data analysis.

\section{Data analysis and results}\label{sec:analysis}
\begin{figure}
\centering
\includegraphics[width=\hsize]{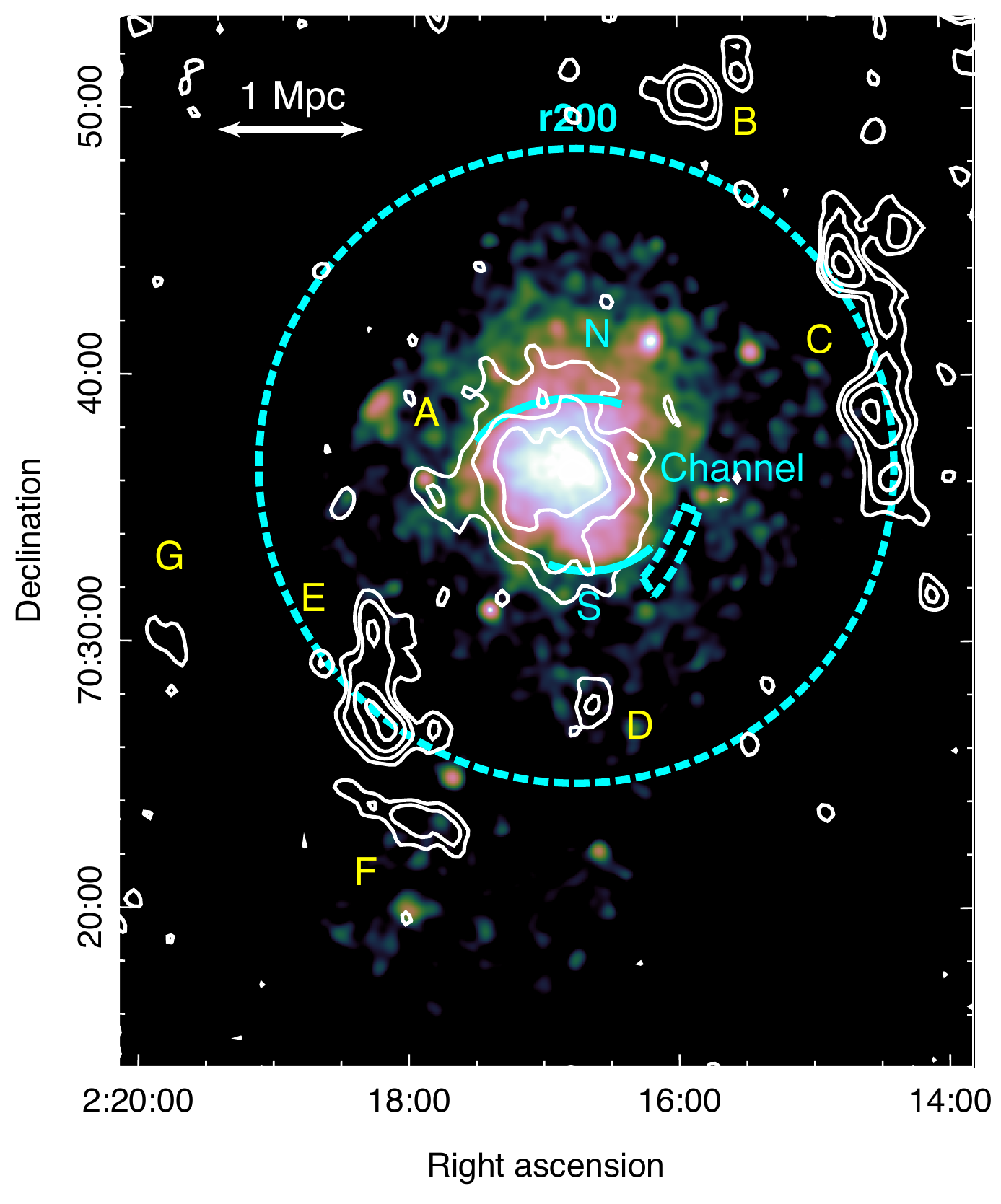}
\caption{Multi-wavelength view of the merging cluster ClG 0217+70. The color image is the NXB subtracted, vignetting corrected, and adaptively smoothed 1--3 keV \emph{Chandra} flux map. The white contours are compact source subtracted VLA $L$-band D configuration radio intensities at $3\sigma_\mathrm{rms}\times[1, 2, 4, 8]$ levels, where $\sigma_\mathrm{rms}=70\ \mu\mathrm{Jy}$. Individual diffuse radio sources are labeled following the terms of \citet{2011ApJ...727L..25B}, where source A is the giant radio halo and B--G are relic candidates. The dashed circle represents $r_{200}$. The northern and southern X-ray surface brightness discontinuities and the western ``channel'' are indicated by cyan arcs and an annulus sector. }
\label{fig:image}

\end{figure}

For the imaging analysis, we used the task \texttt{fluximage} to extract the 1 -- 3 keV count map and the corresponding exposure map with vignetting correction. The non X-ray background (NXB) map is generated from the stowed background file, which is reprojected to the observation frame by using \texttt{reproject\_events} and is scaled by the 9 -- 12 keV count rate. The exposure map is applied to the count map after the NXB subtraction to produce the flux map (see Fig. \ref{fig:image}). For the spectral analysis, we used the task \texttt{blanksky} to create the tailored blank sky background. Source and background spectra are created using the script \texttt{specextract}, where the weighted redistribution matrix files and ancillary response files are created by \texttt{mkwarf} and \texttt{mkacisrmf}, respectively. The background spectra are scaled by the 9 -- 12 keV count rate. We used SPEX v3.06 \citep{1996uxsa.conf..411K,kaastra_j_s_2018_2419563} to fit the spectra. The reference protosolar element abundance table is from \citet{LandoltBornstein2009:sm_lbs_978-3-540-88055-4_34}. The 0.5--7.0 keV energy range of all spectra were used and optimally binned \citep{2016A&A...587A.151K}. The C-statistics \citep{1979ApJ...228..939C} were adopted as the likelihood function in the fit. We used spectral models $red\times hot\times cie$ to fit the spectra of the ICM, where $red$ represents the cosmological redshift, $cie$ is the emission from hot cluster gas in collisional ionization equilibrium, and we fixed the temperature of the $hot$ model to $5\times10^{-4}$ keV to mimic the absorption from neutral gas in our Galaxy. In the $cie$ model, we coupled the abundances of all metals (\element{Li} to \element{Zn}) with \element{Fe}.

\begin{table*}
\caption{Best-fit parameters and auxiliary information for the spectra extracted from different annuli.} \label{tab:spec}
\centering
\begin{tabular}{ccccccc}
\hline\hline
Region & $z$ & $n_\mathrm{H}$ & $kT$  & $Z$ & S/B & C-stat / d.o.f. \\
&&($10^{21}$ cm$^{-2}$)&(keV)&($Z_\sun$)&& \\
\hline
0''--100'' & $0.190\pm0.010$ & $8.2\pm0.5$ & $10.0\substack{+1.6\\-1.1}$ & $0.52\pm0.15$ & 11.6 & $136.9/126$\\
100''--200'' & $0.184\pm0.011$ & $8.1\pm0.4$ & $7.8\pm0.8$ & $0.35\pm0.11$ & 4.5 & $128.5/126$\\
200''--300'' & $0.174\pm0.009$ & $6.8\pm0.7$ & $6.8\pm1.1$ & $0.60\pm0.21$ & 1.0 & $122.9/124$\\
0''--500'' ($r_{500}$) & $0.180\pm0.006$ & $7.3\pm0.3$ & $8.3\pm0.7$ & $0.43\pm0.09$ & 1.7 & $145.1/142$ \\
\hline

\end{tabular}

\end{table*}

\subsection{Spectral properties and X-ray redshift}\label{sec:spec}

This object is at a low Galactic latitude and has high Galactic absorption. 
The tool \texttt{nhtot}\footnote{\url{https://www.swift.ac.uk/analysis/nhtot/index.php}}, which includes the absorption from both atomic and molecular hydrogen \citep{2013MNRAS.431..394W}, suggests $n_\mathrm{H,total}=4.77\times10^{21}$ cm$^{-2}$. However, fixing the $n_\mathrm{H}$ to this value leads to significant residuals in the soft band.
The best-fit $n_\mathrm{H}$ from three annuli centered at the X-ray peak and having different source-to-background ratios (S/B) are consistent with each other and all imply a Galactic $n_\mathrm{H}$ that is higher than the \texttt{nhtot} database value (see Table \ref{tab:spec}). This suggests the low-energy residuals are not due to incorrect modeling of the Galactic Halo foreground. 

Allowing $n_\mathrm{H}$ as a free parameter, the temperature in the three central annuli considered for this analysis is $kT\sim8$~keV. The $kT-r_{200}$ scaling relation of \citet{2009ApJ...691.1307H} then implies $r_{200}=2.3$ Mpc. We also note that the  best-fit redshift for all three annuli is $z\sim0.18$, which is much higher then the previous estimation $z=0.0655$ \citep{2011ApJ...727L..25B}. 

\begin{figure}
\centering
\includegraphics[width=\hsize]{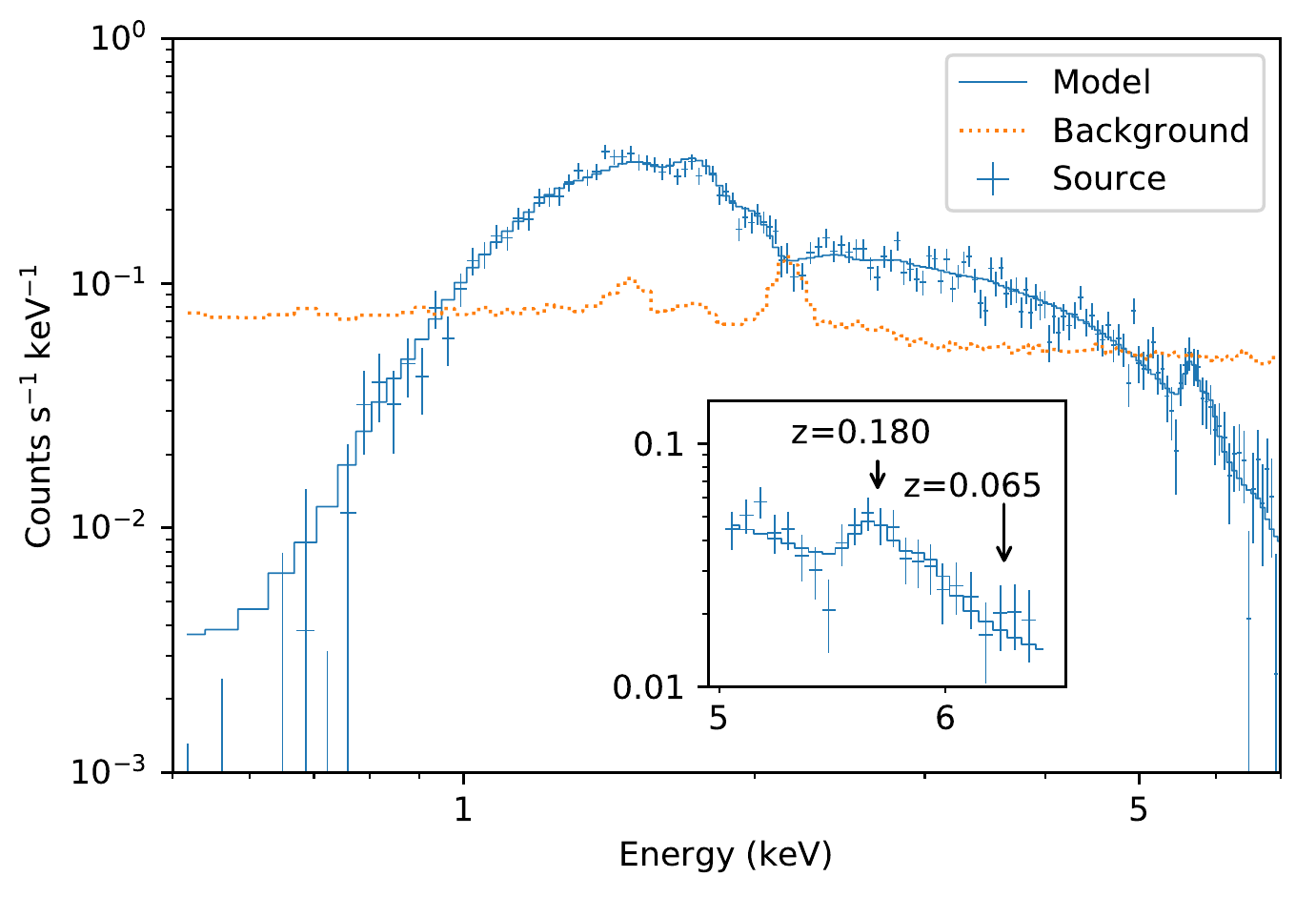}
\caption{Spectrum inside $r_{500}$. The dotted line is the blank sky background that is subtracted. The box is a zoomed-in view of the bump that consists of \ion{Fe}{XXV} He$\alpha$ and \ion{Fe}{XXVI} Ly$\alpha$ lines. The positions of the \ion{Fe}{XXV} He$\alpha$ line at different redshifts are indicated. }
\label{fig:spec}

\end{figure}

To confirm these findings we further extract a spectrum from the cluster's central $r_{500}$ region, which we estimate as $r_{500}\approx0.65r_{200}=1.47$ Mpc \citep{2013SSRv..177..195R}. For our assumed cosmology, this corresponds to $\sim500$'' at $z\sim0.18$. The best-fit redshift within this aperture is indeed $z=0.180\pm0.006$, and the Fe emission lines are clearly visible in the X-ray spectrum (see Fig. \ref{fig:spec}). Other parameters are listed in the fourth row of Table \ref{tab:spec}. 
\citet{2011A&A...529A..65Y} demonstrates that for X-ray CCD spectra, in the condition of $\Delta C_\mathrm{stat}\equiv C_{\mathrm{stat},Z=0}-C_\mathrm{stat, best-fit}>9$, the X-ray redshifts closely agree with the optical spectroscopic redshifts, and the value of our data $\Delta C_\mathrm{stat}=20$ corresponds to an accuracy of $\sigma_z=0.016$. We note that the WHL galaxy cluster catalog \citep{2012ApJS..199...34W}, compiled based on Sloan Digital Sky Survey (SDSS) -III photometric redshifts,  contains a source WHL J021648.6+703646 which overlaps spatially with ClG 0217+70. The brightest cluster galaxy (BCG) of this source is located $20\arcsec$ from the X-ray peak and has $z=0.24\pm0.05$. Therefore, although the quality of the SDSS photometric redshift is poor due to the high Galactic extinction, it is consistent with the presence of a cluster at $z=0.18$. 

With the updated redshift, the largest linear size (LLS) of the radio halo detected by the Very Large Array (VLA) \citep{2011ApJ...727L..25B} reaches 1.6 Mpc, making this cluster the sixth largest known radio halo (see \citealt{2012A&ARv..20...54F} for comparison). The LLS of the western relic candidate (source C in Fig. \ref{fig:image}) reaches 2.3 Mpc, becoming the second largest among the radio relics detected to date (see {\citealt{2014MNRAS.444.3130D} for comparison). Recent Low Frequency Array (LOFAR) data shows that this relic candidate extends even farther to a size of 2.9 Mpc at 150 MHz (Hoang et al. in prep.). 
Additionally, if we assume the X-ray peak as the cluster center, the projected distance $D_\mathrm{projected,rc}$ of the western relic candidate is 2.2 Mpc, which is the second largest among the currently known sample. The easternmost candidate (source G in Fig. \ref{fig:image}), whose distance reaches 2.9 Mpc, becomes the record holder for the farthest radio relic known with respect to the center of any galaxy cluster, surpassing the southeastern relic in PSZ1 G287.00+32.90 \citep{2014ApJ...785....1B} with $D_\mathrm{projected,rc}=2.8$ Mpc. 

Importantly, all relic candidates have a $D_\mathrm{projected}>r_{500}$. Except source D, all candidates are located at $\gtrsim r_{200}$ (see Fig. \ref{fig:image}). This cluster may thus provide a good example of runaway merging shocks \citep{2019MNRAS.488.5259Z}, which are long-lived in the habitable zone in the cluster outskirts.

\subsection{X-ray morphology and surface brightness discontinuities}
The X-ray flux map (Fig. \ref{fig:image}) shows a single-peaked and irregular morphology inside $r_{2500}$. The X-ray core is elongated in the NW-SE orientation, and its location matches the peak of the radio halo. 
Unlike some other typical binary on-axis merging systems with double relics, for example  Abell 3376 \citep{2006Sci...314..791B}, ZwCl 0008.8+5215 \citep{2019ApJ...873...64D}, and El Gordo \citep{2012ApJ...748....7M}, the morphology of this cluster does not show an outbound subcluster, perhaps indicating a later merger phase. 

Previous work shows that the presence of a radio halo is related to the dynamical state of a cluster, which can be quantified by X-ray morphological parameters \citep{2010ApJ...721L..82C}. Following the methods in \citet{2010ApJ...721L..82C}, we calculate the centroid shift $w$ \citep{1993ApJ...413..492M,2006MNRAS.373..881P} and the concentration parameter $c$ \citep{2008A&A...483...35S}. The result $(w,c)=(0.046,0.11)$ is located in the quadrant where most of the clusters host a radio halo (see Fig. 1 in \citealt{2010ApJ...721L..82C}). 

\begin{figure*}[]
\centering
\begin{tabular}{ccc}

\includegraphics[width=0.3\hsize]{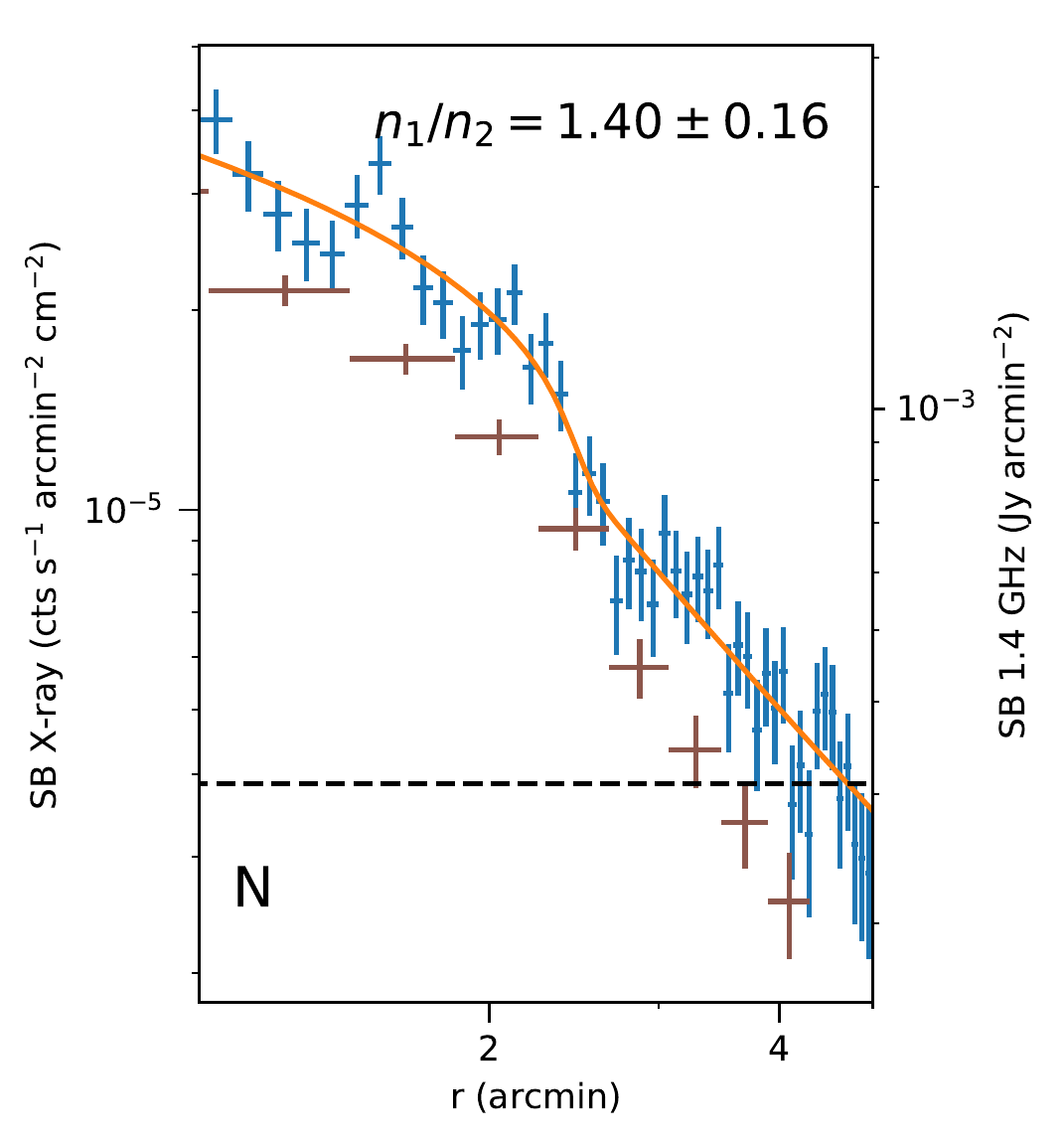} &
\includegraphics[width=0.3\hsize]{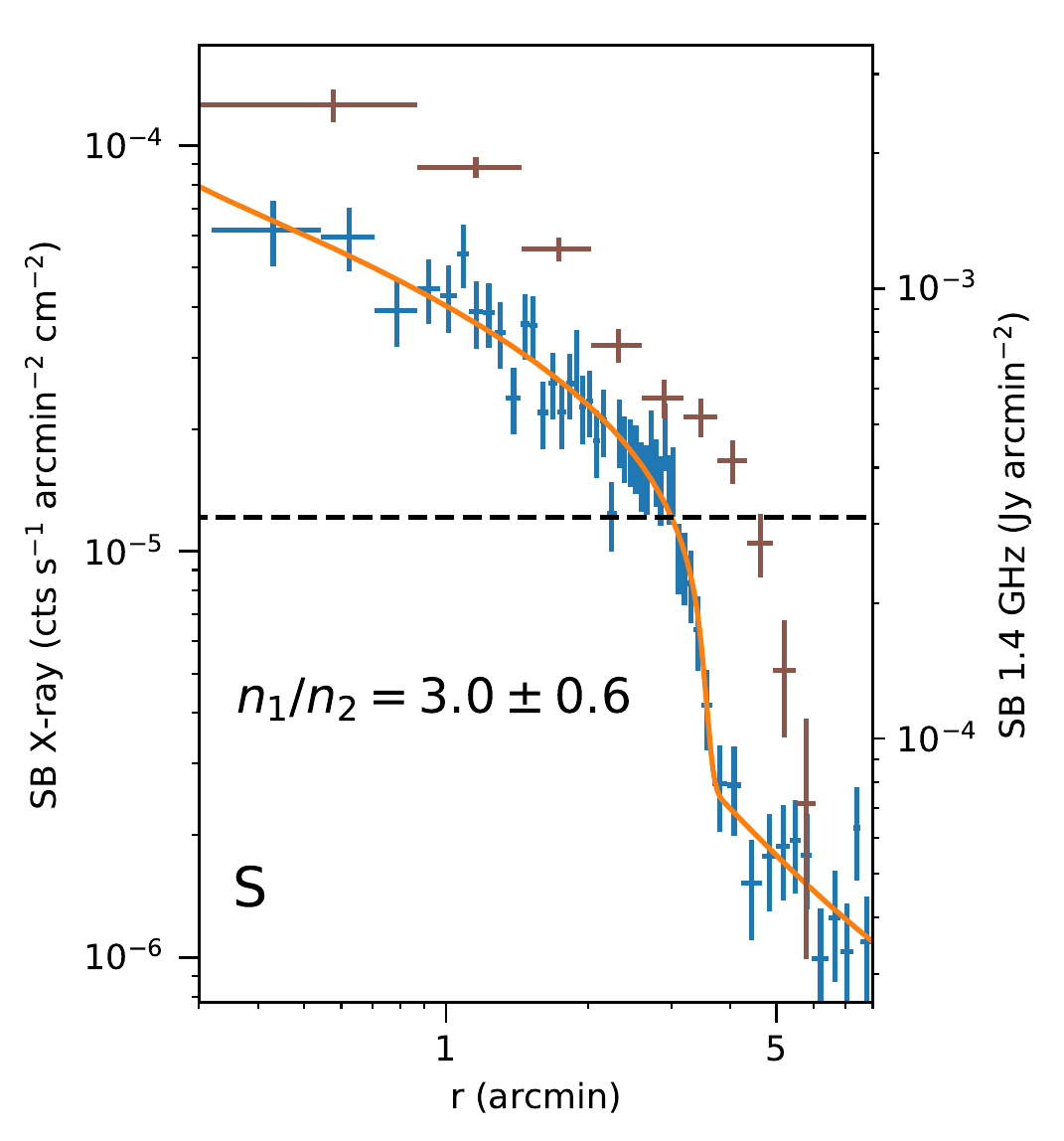} &

\includegraphics[width=0.3\hsize]{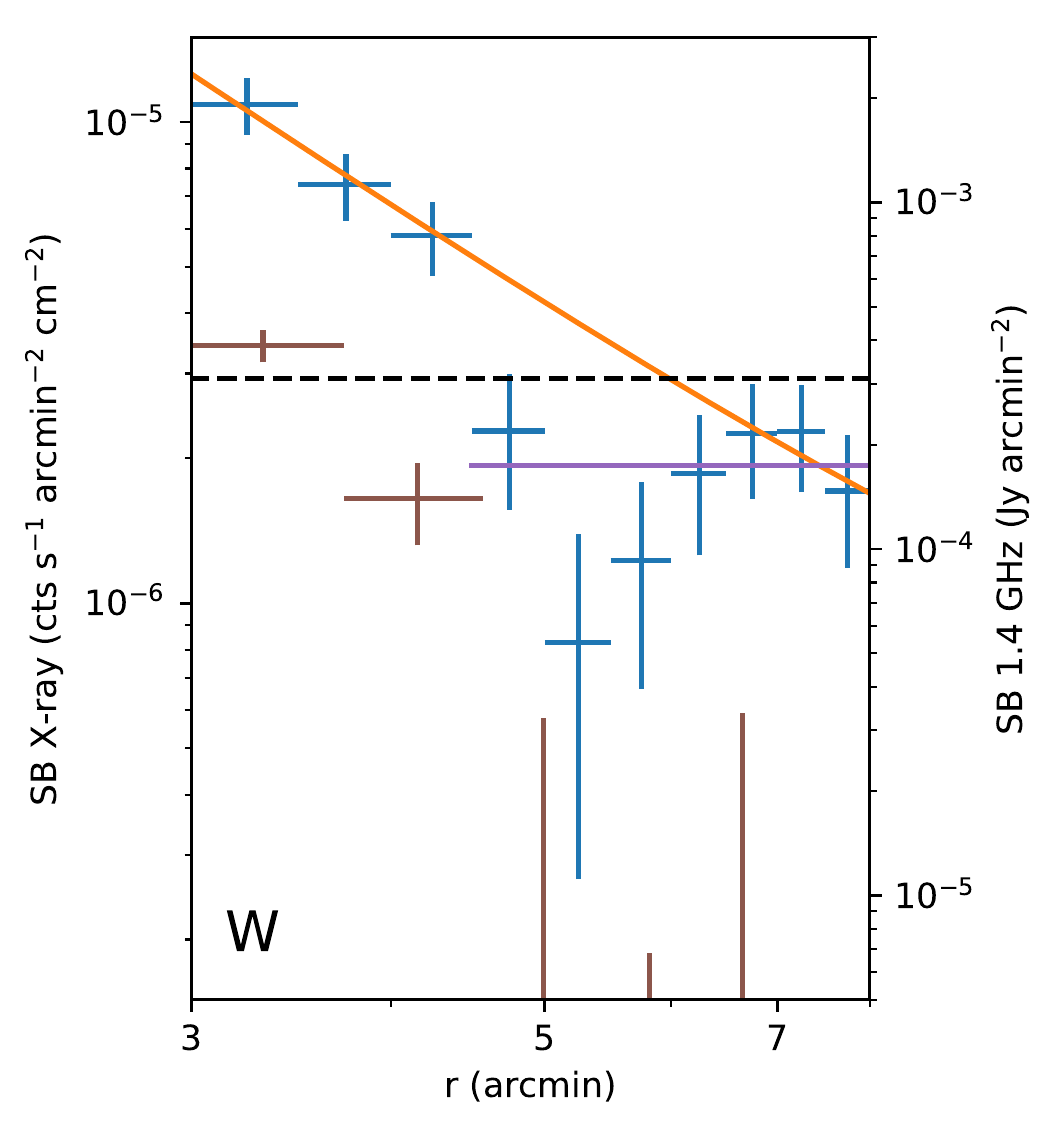}
\end{tabular}
\caption{
X-ray surface brightness profiles (blue) and the best-fit model (orange) in each extraction region. The radio surface brightness profiles are plotted as brown points. The horizontal dashed line is the $3\times\sigma_\mathrm{rms}$ level of the radio map. In the western channel region, In addition to the best-fit power-law model (orange), the constant model after the jump is plotted as the purple line.}
\label{fig:sb}
\end{figure*}

In addition, we find two X-ray surface brightness discontinuities about 460 kpc and 680 kpc toward the north and south of the core (see Fig. \ref{fig:image}). We extract and fit the surface brightness profiles assuming an underlying spherically projected double power-law  density model \citep{2009ApJ...704.1349O}.
To account for a possible mismatch between the extraction regions and the actual curvature of the edge, we smooth the projected model profile with a $\sigma=0.1\arcmin$ Gaussian kernel.
The cosmic X-ray background is modeled as a constant $S_\mathrm{bg}=4\times10^{-7}$ count s$^{-1}$ arcmin$^{-2}$ cm$^{-2}$, obtained by fitting the azimuthally averaged radial surface brightness profile using a double $\beta$-model \citep{1976A&A....49..137C} plus a constant model.

The profiles and best-fit models are plotted in Fig. \ref{fig:sb}. The best-fit density jump of the southern and northern edges are $C=3.0\pm0.6$ and $C=1.40\pm0.16$, respectively. 
We also overplot the VLA $L$-band D configuration \citep{2011ApJ...727L..25B} radio surface brightness profiles. In the northern region the slope of the radio surface brightness profile changes from $0.85\pm0.05$ to $2.4\pm0.3$ at the radius of 400 kpc.
The southern region shows a marked X-ray jump that is, however, misaligned with a steep drop-off in the radio brightness profile, which occurs $\sim 2\arcmin = 370$ kpc  farther out. This sharp drop is unlikely due to the flux loss in interferometric observations. The largest angular scale of D configuration is $\sim16'$. Only emission that is smooth on scales $>10'$ is subject to significant flux loss (on the order of $10\%$ or greater, \citealt{2011ApJ...727L..25B}), while the features discussed here are on much smaller scales. 


The thermodynamic properties of these edges are still unclear due to the short exposure time; however, for the southern edge to be a cold front, the temperature on the faint side would have to be very high, $16$~keV. If either of the two edges is confirmed as a shock front, this cluster could be another rare case where the X-ray shock is associated with the edge of a radio halo, for example  Abell 520 \citep{2019A&A...622A..20H}, the Bullet Cluster \citep{2014MNRAS.440.2901S}, the Coma Cluster \citep{2011MNRAS.412....2B,2013A&A...554A.140P},  and the Toothbrush Cluster \citep{2016ApJ...818..204V}.

Apart from the two surface brightness discontinuities, we also find a large scale X-ray channel in the southwestern part of the cluster (the dashed region in Fig. \ref{fig:image}). The length is at least 700 kpc and the width can be over 200 kpc. We use a single power-law density model to fit this profile, where we ignore the data points from 4.5 to 6 arcmin. The lowest point is below the power-law model by $5.3\sigma$. Alternatively, the large difference between $4\arcmin$ and $5\arcmin$ can be interpreted as a surface brightness edge, with the channel being right outside the edge. In this case, if we use a constant model to fit the data points from 4.5 to 8 arcmin excepting the dip, the lowest point is $2.0\sigma$ below the model. The density in the channel region is $<50\%$ of the power-law model and is $65\%$ of the constant model. The radio surface brightness also decreases sharply before the channel. Inside the channel, the upper limit of the radio emission is $7\times10^{-5}$ Jy arcmin$^{-2}$.

\section{Discussion}\label{sec:discussion}

\subsection{Radio halo scaling relations}
This cluster was believed to be an X-ray underluminous or a radio overluminous source in the radio halo $L_\mathrm{X}-P_\mathrm{1.4\ GHz}$ diagram \citep{2011ApJ...727L..25B}. Meanwhile, using the $M_{500}-L_{X}$ scaling relation \citep{2009A&A...498..361P}, \citet{2017MNRAS.470.3465B} found this object to be an outlier in the $M_{500}-P_\mathrm{1.4\ GHz}$ relation.
The new redshift reported here leads to an updated 1.4 GHz radio power of $(5.33\pm0.08)\times10^{24}$ W Hz$^{-1}$ and an X-ray luminosity inside $r_{500}$ of $L_{0.1-2.4 \mathrm{keV}}=(7.19\pm0.12)\times10^{44}$ erg s$^{-1}$, bringing this cluster into agreement with the expected radio halo $L_\mathrm{X}-P_\mathrm{1.4\ GHz}$ relationship. Furthermore, using the $M_{500}$--$kT$ scaling relation of \citet{2007A&A...474L..37A} we obtain $M_{500}=(1.06\pm0.11)\times10^{15}\ M_\sun$. This cluster then also follows the $M_{500}-P_\mathrm{1.4\ GHz}$ scaling relation \citep{2013ApJ...777..141C} (see Fig. \ref{fig:relation}). The Sunyaev-Zeldovich (SZ) effect of such a massive cluster should be detected by \emph{Planck}. However, due to its low Galactic latitude, it is not in the second \emph{Planck} SZ catalog \citep{2016A&A...594A..27P}.

The remaining two outliers on the $L_\mathrm{X}-P_\mathrm{radio}$ scaling relation are \object{Abell 1213} and \object{Abell 523}.
In Abell 1213, the diffuse emission is on scales of only 200 kpc and is dominated by bright filamentary structures \citep{2009A&A...507.1257G}. Abell 523 has a unique linear structure unlike other radio halos \citep{2011A&A...530L...5G}. With the results presented here, there may thus be no known regular radio halo that does not follow these scaling relations.

\begin{figure}[]
\centering
\includegraphics[width=.9\hsize]{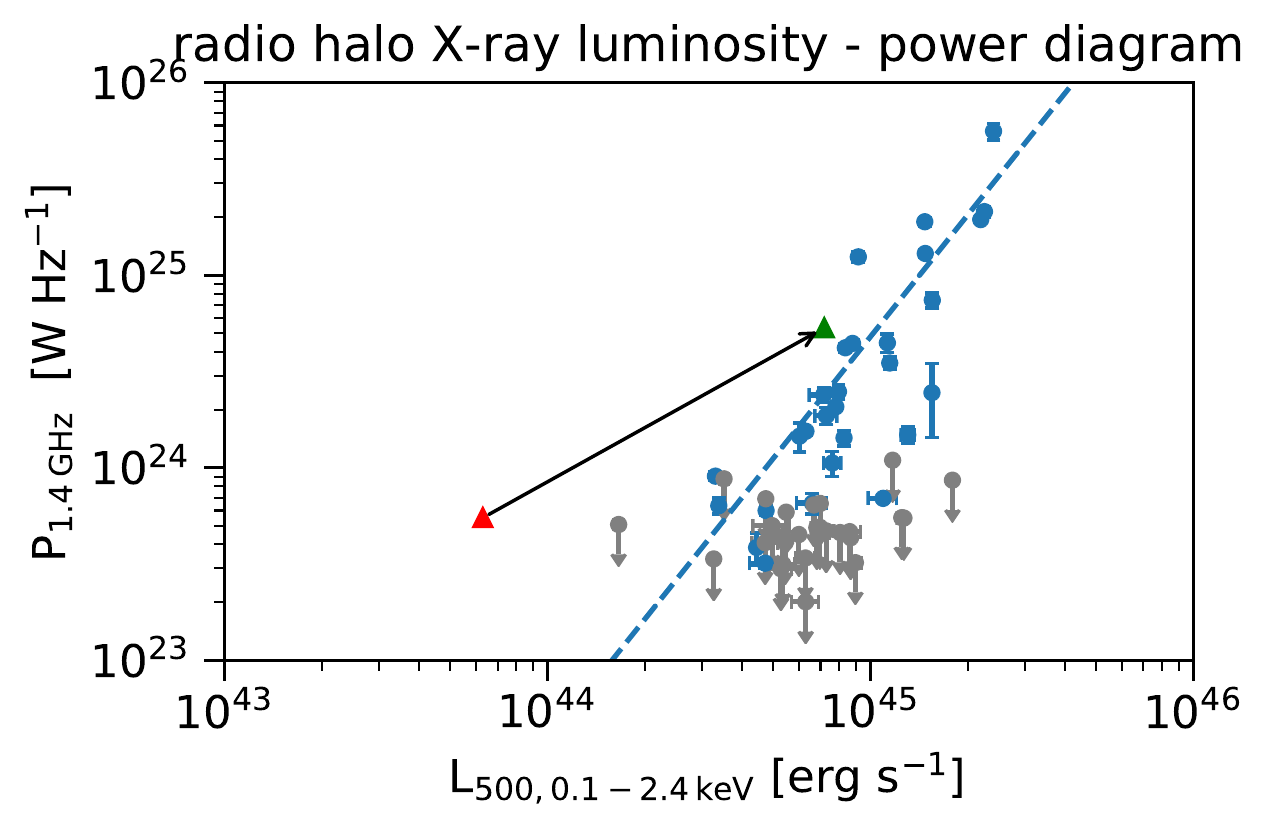} 
\includegraphics[width=.9\hsize]{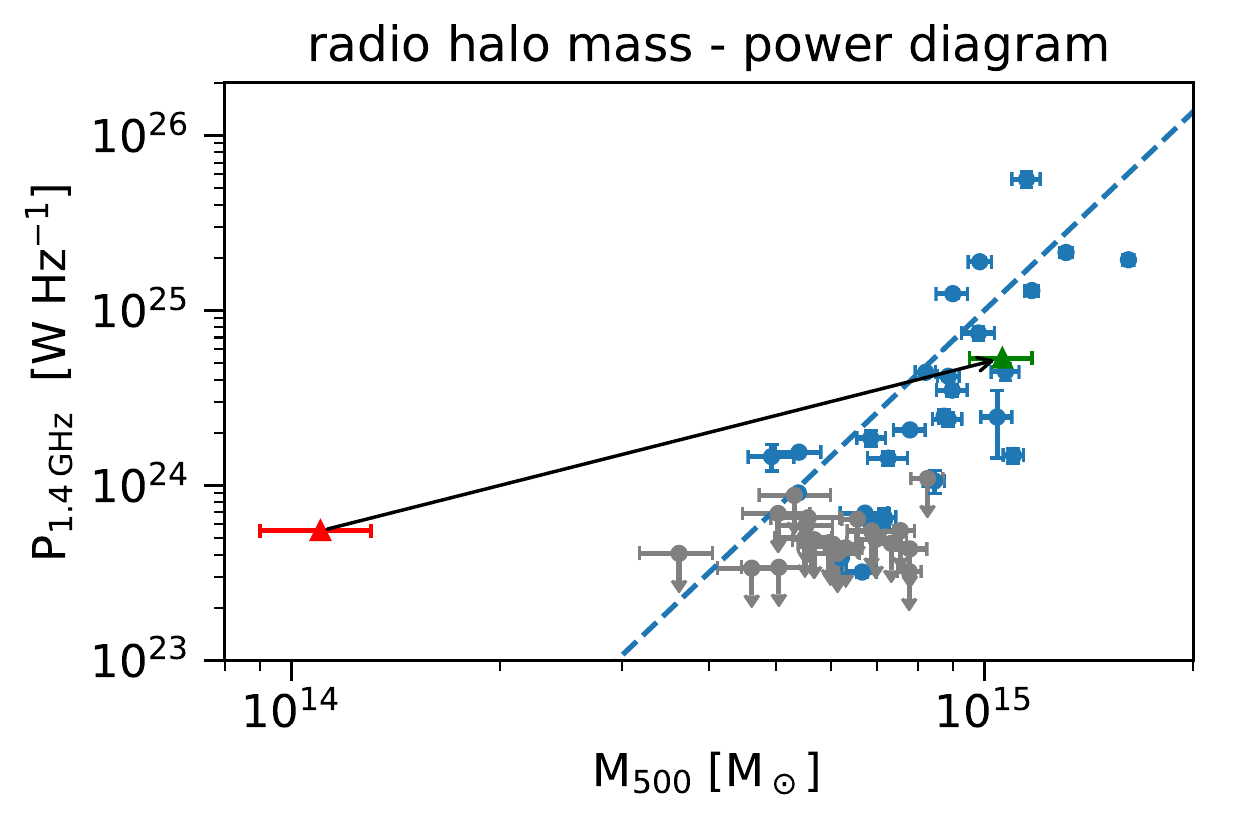} 

\caption{$L_\mathrm{X}-P_\mathrm{1.4\ GHz}$ (top) and $M_{500}-P_\mathrm{1.4\ GHz}$ (bottom) diagrams of radio halos. Figures are modified based on the version in \citet{2019SSRv..215...16V}, where the data samples are from \citet{2013ApJ...777..141C}, \citet{2015A&A...579A..92K} and \citet{2018A&A...609A..61C}. Blue points are clusters with radio halos and gray upper limits are clusters without radio halos. Blue dashed lines are the best-fit scaling relations from \citet{2013ApJ...777..141C}. The red triangle is ClG 0217+70 using the previous redshift, the green triangle is the corrected value. The error bars of $L_\mathrm{X}$ and $P_\mathrm{1.4\ GHz}$ of ClG 0217+70 are smaller than the plot symbols.}
\label{fig:relation}
\end{figure}

\subsection{Western X-ray channel}
We observe an X-ray deficit in the western part of the cluster, where the gas density is about half of that in the inner region. This channel-like structure can be a compressed heated region between the main cluster and an in-falling group, as has been seen in  Abell 85, for example \citep{2015MNRAS.448.2971I}, or between colliding subclusters \citep[e.g., \object{Abell 521;}][]{2013ApJ...764...82B}. Although no significant X-ray substructure is seen outside the channel in ClG 0217+70, an infalling group that has been stripped of its gas content early during the merger could still produce the observed feature.

Alternatively, non-thermal pressure, either in the form of turbulent motions or enhanced magnetic fields that push out the thermal gas, may play a role. This explanation has been proposed for similar substructures observed along a cold front in the Virgo Cluster \citep{2016MNRAS.455..846W}, and in \object{Abell 520} \citep{2016ApJ...833...99W} and \object{Abell 2142} \citep{2018ApJ...868...45W}. 
Assuming the ICM is isothermal across the channel, to compensate the pressure deficit of  at least 35\%, the non-thermal pressure should be equal to the thermal pressure. If the magnetic field enhancement is alone responsible for this non-thermal pressure, $B\sim15$ $\mu$G is required and the corresponding thermal-to-magnetic pressure ratio reaches $\beta\sim2$. Such a considerable local magnetic field enhancement is indeed seen in MHD simulations of sloshing cold fronts \citep{2011ApJ...743...16Z}. One might expect that this high magnetic field would lead to a detectable level of radio emission in the channel, which is disfavored by the current VLA observation. However, since the exact underlying relativistic electron distribution is unknown, this scenario cannot be ruled out. If on the other hand the non-thermal pressure is entirely due to turbulent motions, the turbulent Mach number $\mathcal{M}_\mathrm{turb}=\sqrt{2/\gamma\times(\epsilon_\mathrm{turb}/\epsilon_\mathrm{therm})}$ \citep{2009MNRAS.398...23W} should be close to one, which is also very unlikely. In reality of course a combination of all the above factors is also possible.

\subsection{Possible merging scenario} 

The western relic candidate C as well as the eastern relic candidates E, F, and G are likely to be accelerated by two spherical shocks that are centrally symmetric and moving towards NW and SE, respectively. 
The two spherical shocks are presumably created by the first core passage and then propagate to the outskirts of this system. The two discovered surface brightness jumps are in the N-S orientation, which is almost perpendicular to the previous merging axis. Additionally, the projected distance between the cluster center and the two surface brightness discontinuities is $\la700$ kpc, which is much less than the $D_\mathrm{projected}$ of the outermost relics. Both the orientation and the short $D_\mathrm{projected}$ of the surface brightness discontinuities suggest that they are not related to the first core passage event. A possible explanation of this merger might be that it starts as an off-axis merger, after which the two (or more) dark matter halos, as well as the ionized gas, move back to the centroid of the system. The collision of the ICM produces new shocks or cold fronts, but the orientation is different from the first passage. 


\section{Conclusion}\label{sec:conclusion}
We analyzed the 25 ks archival \emph{Chandra} data of the merging galaxy cluster ClG 0217+70. The \emph{Chandra} X-ray data allow us to measure the redshift of the system, which is $z=0.180\pm0.006$. With the updated redshift, the projected physical sizes of the radio halo and radio relic candidates make them some of the largest sources ever discovered. Most of the radio relic candidates have projected distances $\gtrsim r_{200}$.
We measure the averaged temperature inside $r_{500}$ as $kT_{500}=8.3\pm0.6$ keV. 
Using the $kT-M_{500}$ scaling relation, we estimate $M_{500}=(1.06\pm0.11)\times10^{15}\ M_\sun$. The centroid shift $w$ and the concentration parameter $c$ show that the ICM is still dynamically disturbed. Two surface brightness discontinuities are detected with density jumps of $1.40\pm0.16$ in the north and $3.0\pm0.6$ in the south. The southern edge has one of the largest density jumps ever detected in galaxy clusters. We also find a 700 kpc long and $>200$ kpc wide surface brightness channel in the western part of the cluster, which may be indicative of significant compressed heated gas or non-thermal pressure from magnetic fields or turbulence.


In this work, X-ray spectroscopy shows its power of measuring the ICM redshift directly. Its strength will be remarkably exploited in future missions with X-ray microcalorimeters, for example  \emph{XRISM} and \emph{Athena}. 

\begin{acknowledgements}
We thank the anonymous referee for helpful suggestions. The reproduction package of this research is openly available from Zenodo at \href{https://doi.org/10.5281/zenodo.4032424}{10.5281/zenodo.4032424}.
X.Z. acknowledges support from China Scholarship Council. SRON is supported financially by NWO, The Netherlands Organisation for Scientific Research. D.N.H. and C.S. acknowledges support from the ERC-StG DRANOEL, n. 714245. R.J.vW. acknowledges support from the VIDI research programme with project number 639.042.729, which is financed by the Netherlands Organisation for Scientific Research (NWO). Partial support for L.R. comes from US National Science Foundation grant AST17-14205 to the University of Minnesota. This research has made use of data obtained from the Chandra Data Archive and the Chandra Source Catalog, and software provided by the Chandra X-ray Center (CXC) in the application package CIAO.
\end{acknowledgements}

\bibliography{0217}

\begin{thebibliography}{51}
\expandafter\ifx\csname natexlab\endcsname\relax\def\natexlab#1{#1}\fi

\bibitem[{{Arnaud} {et~al.}(2007){Arnaud}, {Pointecouteau}, \&
  {Pratt}}]{2007A&A...474L..37A}
{Arnaud}, M., {Pointecouteau}, E., \& {Pratt}, G.~W. 2007, \aap, 474, L37

\bibitem[{{Bagchi} {et~al.}(2006){Bagchi}, {Durret}, {Neto}, \&
  {Paul}}]{2006Sci...314..791B}
{Bagchi}, J., {Durret}, F., {Neto}, G. B.~L., \& {Paul}, S. 2006, Science, 314,
  791

\bibitem[{{Bonafede} {et~al.}(2017){Bonafede}, {Cassano}, {Br{\"u}ggen},
  {Ogrean}, {Riseley}, {Cuciti}, {de Gasperin}, {Golovich}, {Kale}, {Venturi},
  {van Weeren}, {Wik}, \& {Wittman}}]{2017MNRAS.470.3465B}
{Bonafede}, A., {Cassano}, R., {Br{\"u}ggen}, M., {et~al.} 2017, \mnras, 470,
  3465

\bibitem[{{Bonafede} {et~al.}(2014){Bonafede}, {Intema}, {Br{\"u}ggen},
  {Girardi}, {Nonino}, {Kantharia}, {van Weeren}, \&
  {R{\"o}ttgering}}]{2014ApJ...785....1B}
{Bonafede}, A., {Intema}, H.~T., {Br{\"u}ggen}, M., {et~al.} 2014, \apj, 785, 1

\bibitem[{{Bourdin} {et~al.}(2013){Bourdin}, {Mazzotta}, {Markevitch},
  {Giacintucci}, \& {Brunetti}}]{2013ApJ...764...82B}
{Bourdin}, H., {Mazzotta}, P., {Markevitch}, M., {Giacintucci}, S., \&
  {Brunetti}, G. 2013, \apj, 764, 82

\bibitem[{{Brown} {et~al.}(2011){Brown}, {Duesterhoeft}, \&
  {Rudnick}}]{2011ApJ...727L..25B}
{Brown}, S., {Duesterhoeft}, J., \& {Rudnick}, L. 2011, \apjl, 727, L25

\bibitem[{{Brown} \& {Rudnick}(2011)}]{2011MNRAS.412....2B}
{Brown}, S. \& {Rudnick}, L. 2011, \mnras, 412, 2

\bibitem[{{Brunetti} {et~al.}(2009){Brunetti}, {Cassano}, {Dolag}, \&
  {Setti}}]{2009A&A...507..661B}
{Brunetti}, G., {Cassano}, R., {Dolag}, K., \& {Setti}, G. 2009, \aap, 507, 661

\bibitem[{{Brunetti} \& {Jones}(2014)}]{2014IJMPD..2330007B}
{Brunetti}, G. \& {Jones}, T.~W. 2014, International Journal of Modern Physics
  D, 23, 1430007

\bibitem[{{Cash}(1979)}]{1979ApJ...228..939C}
{Cash}, W. 1979, \apj, 228, 939

\bibitem[{{Cassano} {et~al.}(2013){Cassano}, {Ettori}, {Brunetti},
  {Giacintucci}, {Pratt}, {Venturi}, {Kale}, {Dolag}, \&
  {Markevitch}}]{2013ApJ...777..141C}
{Cassano}, R., {Ettori}, S., {Brunetti}, G., {et~al.} 2013, \apj, 777, 141

\bibitem[{{Cassano} {et~al.}(2010){Cassano}, {Ettori}, {Giacintucci},
  {Brunetti}, {Markevitch}, {Venturi}, \& {Gitti}}]{2010ApJ...721L..82C}
{Cassano}, R., {Ettori}, S., {Giacintucci}, S., {et~al.} 2010, \apjl, 721, L82

\bibitem[{{Cavaliere} \& {Fusco-Femiano}(1976)}]{1976A&A....49..137C}
{Cavaliere}, A. \& {Fusco-Femiano}, R. 1976, \aap, 500, 95

\bibitem[{{Cuciti} {et~al.}(2018){Cuciti}, {Brunetti}, {van Weeren},
  {Bonafede}, {Dallacasa}, {Cassano}, {Venturi}, \&
  {Kale}}]{2018A&A...609A..61C}
{Cuciti}, V., {Brunetti}, G., {van Weeren}, R., {et~al.} 2018, \aap, 609, A61

\bibitem[{{de Gasperin} {et~al.}(2014){de Gasperin}, {van Weeren},
  {Br{\"u}ggen}, {Vazza}, {Bonafede}, \& {Intema}}]{2014MNRAS.444.3130D}
{de Gasperin}, F., {van Weeren}, R.~J., {Br{\"u}ggen}, M., {et~al.} 2014,
  \mnras, 444, 3130

\bibitem[{{Di Gennaro} {et~al.}(2019){Di Gennaro}, {van Weeren},
  {Andrade-Santos}, {Akamatsu}, {Randall}, {Forman}, {Kraft}, {Brunetti},
  {Dawson}, {Golovich}, \& {Jones}}]{2019ApJ...873...64D}
{Di Gennaro}, G., {van Weeren}, R.~J., {Andrade-Santos}, F., {et~al.} 2019,
  \apj, 873, 64

\bibitem[{{Feretti} {et~al.}(2012){Feretti}, {Giovannini}, {Govoni}, \&
  {Murgia}}]{2012A&ARv..20...54F}
{Feretti}, L., {Giovannini}, G., {Govoni}, F., \& {Murgia}, M. 2012, \aapr, 20,
  54

\bibitem[{{Giovannini} {et~al.}(2009){Giovannini}, {Bonafede}, {Feretti},
  {Govoni}, {Murgia}, {Ferrari}, \& {Monti}}]{2009A&A...507.1257G}
{Giovannini}, G., {Bonafede}, A., {Feretti}, L., {et~al.} 2009, \aap, 507, 1257

\bibitem[{{Giovannini} {et~al.}(2011){Giovannini}, {Feretti}, {Girardi},
  {Govoni}, {Murgia}, {Vacca}, \& {Bagchi}}]{2011A&A...530L...5G}
{Giovannini}, G., {Feretti}, L., {Girardi}, M., {et~al.} 2011, \aap, 530, L5

\bibitem[{{Hales} {et~al.}(1995){Hales}, {Waldram}, {Rees}, \&
  {Warner}}]{1995MNRAS.274..447H}
{Hales}, S.~E.~G., {Waldram}, E.~M., {Rees}, N., \& {Warner}, P.~J. 1995,
  \mnras, 274, 447

\bibitem[{{Henry} {et~al.}(2009){Henry}, {Evrard}, {Hoekstra}, {Babul}, \&
  {Mahdavi}}]{2009ApJ...691.1307H}
{Henry}, J.~P., {Evrard}, A.~E., {Hoekstra}, H., {Babul}, A., \& {Mahdavi}, A.
  2009, \apj, 691, 1307

\bibitem[{{Hoang} {et~al.}(2019){Hoang}, {Shimwell}, {van Weeren}, {Brunetti},
  {R{\"o}ttgering}, {Andrade-Santos}, {Botteon}, {Br{\"u}ggen}, {Cassano},
  {Drabent}, {de Gasperin}, {Hoeft}, {Intema}, {Rafferty}, {Shweta}, \&
  {Stroe}}]{2019A&A...622A..20H}
{Hoang}, D.~N., {Shimwell}, T.~W., {van Weeren}, R.~J., {et~al.} 2019, \aap,
  622, A20

\bibitem[{{Ichinohe} {et~al.}(2015){Ichinohe}, {Werner}, {Simionescu}, {Allen},
  {Canning}, {Ehlert}, {Mernier}, \& {Takahashi}}]{2015MNRAS.448.2971I}
{Ichinohe}, Y., {Werner}, N., {Simionescu}, A., {et~al.} 2015, \mnras, 448,
  2971

\bibitem[{{Kaastra} \& {Bleeker}(2016)}]{2016A&A...587A.151K}
{Kaastra}, J.~S. \& {Bleeker}, J.~A.~M. 2016, \aap, 587, A151

\bibitem[{{Kaastra} {et~al.}(1996){Kaastra}, {Mewe}, \&
  {Nieuwenhuijzen}}]{1996uxsa.conf..411K}
{Kaastra}, J.~S., {Mewe}, R., \& {Nieuwenhuijzen}, H. 1996, in UV and X-ray
  Spectroscopy of Astrophysical and Laboratory Plasmas, 411--414

\bibitem[{Kaastra {et~al.}(2018)Kaastra, Raassen, de~Plaa, \&
  Gu}]{kaastra_j_s_2018_2419563}
Kaastra, J.~S., Raassen, A. J.~J., de~Plaa, J., \& Gu, L. 2018, SPEX X-ray
  spectral fitting package. Zenodo. https://doi.org/10.5281/zenodo.2419563

\bibitem[{{Kale} {et~al.}(2015){Kale}, {Venturi}, {Giacintucci}, {Dallacasa},
  {Cassano}, {Brunetti}, {Cuciti}, {Macario}, \&
  {Athreya}}]{2015A&A...579A..92K}
{Kale}, R., {Venturi}, T., {Giacintucci}, S., {et~al.} 2015, \aap, 579, A92

\bibitem[{Lodders {et~al.}(2009)Lodders, Palme, \&
  Gail}]{LandoltBornstein2009:sm_lbs_978-3-540-88055-4_34}
Lodders, K., Palme, H., \& Gail, H.-P. 2009, 4.4 Abundances of the elements in
  the Solar System: Datasheet from Landolt-B{\"o}rnstein - Group VI Astronomy
  and Astrophysics {\textperiodcentered} Volume 4B: ``Solar System'' in
  SpringerMaterials (https://doi.org/10.1007/978-3-540-88055-4{\_}34)

\bibitem[{{Menanteau} {et~al.}(2012){Menanteau}, {Hughes}, {Sif{\'o}n},
  {Hilton}, {Gonz{\'a}lez}, {Infante}, {Barrientos}, {Baker}, {Bond}, {Das},
  {Devlin}, {Dunkley}, {Hajian}, {Hincks}, {Kosowsky}, {Marsden}, {Marriage},
  {Moodley}, {Niemack}, {Nolta}, {Page}, {Reese}, {Sehgal}, {Sievers},
  {Spergel}, {Staggs}, \& {Wollack}}]{2012ApJ...748....7M}
{Menanteau}, F., {Hughes}, J.~P., {Sif{\'o}n}, C., {et~al.} 2012, \apj, 748, 7

\bibitem[{{Mohr} {et~al.}(1993){Mohr}, {Fabricant}, \&
  {Geller}}]{1993ApJ...413..492M}
{Mohr}, J.~J., {Fabricant}, D.~G., \& {Geller}, M.~J. 1993, \apj, 413, 492

\bibitem[{{Owers} {et~al.}(2009){Owers}, {Nulsen}, {Couch}, \&
  {Markevitch}}]{2009ApJ...704.1349O}
{Owers}, M.~S., {Nulsen}, P. E.~J., {Couch}, W.~J., \& {Markevitch}, M. 2009,
  \apj, 704, 1349

\bibitem[{{Planck Collaboration} {et~al.}(2013){Planck Collaboration}, {Ade},
  {Aghanim}, {Arnaud}, {Ashdown}, {Atrio-Barandela}, {Aumont}, {Baccigalupi},
  {Balbi}, {Banday}, {Barreiro}, {Bartlett}, {Battaner}, {Benabed},
  {Beno{\^\i}t}, {Bernard}, {Bersanelli}, {Bikmaev}, {B{\"o}hringer},
  {Bonaldi}, {Bond}, {Borrill}, {Bouchet}, {Bourdin}, {Brown}, {Brown},
  {Burenin}, {Burigana}, {Cabella}, {Cardoso}, {Carvalho}, {Catalano},
  {Cay{\'o}n}, {Chiang}, {Chon}, {Christensen}, {Churazov}, {Clements},
  {Colafrancesco}, {Colombo}, {Coulais}, {Crill}, {Cuttaia}, {Da Silva},
  {Dahle}, {Danese}, {Davis}, {de Bernardis}, {de Gasperis}, {de Rosa}, {de
  Zotti}, {Delabrouille}, {D{\'e}mocl{\`e}s}, {D{\'e}sert}, {Dickinson},
  {Diego}, {Dolag}, {Dole}, {Donzelli}, {Dor{\'e}}, {D{\"o}rl}, {Douspis},
  {Dupac}, {En{\ss}lin}, {Eriksen}, {Finelli}, {Flores-Cacho}, {Forni},
  {Frailis}, {Franceschi}, {Frommert}, {Galeotta}, {Ganga},
  {G{\'e}nova-Santos}, {Giard}, {Gilfanov}, {Gonz{\'a}lez-Nuevo}, {G{\'o}rski},
  {Gregorio}, {Gruppuso}, {Hansen}, {Harrison}, {Henrot-Versill{\'e}},
  {Hern{\'a}ndez-Monteagudo}, {Hildebrandt}, {Hivon}, {Hobson}, {Holmes},
  {Hornstrup}, {Hovest}, {Huffenberger}, {Hurier}, {Jaffe}, {Jagemann},
  {Jones}, {Juvela}, {Keih{\"a}nen}, {Khamitov}, {Kneissl}, {Knoche}, {Knox},
  {Kunz}, {Kurki-Suonio}, {Lagache}, {L{\"a}hteenm{\"a}ki}, {Lamarre},
  {Lasenby}, {Lawrence}, {Le Jeune}, {Leonardi}, {Lilje}, {Linden-V{\o}rnle},
  {L{\'o}pez-Caniego}, {Lubin}, {Mac{\'\i}as-P{\'e}rez}, {Maffei}, {Maino},
  {Mand olesi}, {Maris}, {Marleau}, {Mart{\'\i}nez-Gonz{\'a}lez}, {Masi},
  {Massardi}, {Matarrese}, {Matthai}, {Mazzotta}, {Mei}, {Melchiorri}, {Melin},
  {Mendes}, {Mennella}, {Mitra}, {Miville-Desch{\^e}nes}, {Moneti}, {Montier},
  {Morgante}, {Munshi}, {Murphy}, {Naselsky}, {Natoli}, {N{\o}rgaard-Nielsen},
  {Noviello}, {Novikov}, {Novikov}, {Osborne}, {Pajot}, {Paoletti},
  {Perdereau}, {Perrotta}, {Piacentini}, {Piat}, {Pierpaoli}, {Piffaretti},
  {Plaszczynski}, {Pointecouteau}, {Polenta}, {Ponthieu}, {Popa}, {Poutanen},
  {Pratt}, {Prunet}, {Puget}, {Rachen}, {Rebolo}, {Reinecke}, {Remazeilles},
  {Renault}, {Ricciardi}, {Riller}, {Ristorcelli}, {Rocha}, {Roman}, {Rosset},
  {Rossetti}, {Rubi{\~n}o-Mart{\'\i}n}, {Rudnick}, {Rusholme}, {Sandri},
  {Savini}, {Schaefer}, {Scott}, {Smoot}, {Stivoli}, {Sudiwala}, {Sunyaev},
  {Sutton}, {Suur-Uski}, {Sygnet}, {Tauber}, {Terenzi}, {Toffolatti}, {Tomasi},
  {Tristram}, {Tuovinen}, {T{\"u}rler}, {Umana}, {Valenziano}, {Van Tent},
  {Varis}, {Vielva}, {Villa}, {Vittorio}, {Wade}, {Wandelt}, {Welikala},
  {White}, {Yvon}, {Zacchei}, {Zaroubi}, \& {Zonca}}]{2013A&A...554A.140P}
{Planck Collaboration}, {Ade}, P.~A.~R., {Aghanim}, N., {et~al.} 2013, \aap,
  554, A140

\bibitem[{{Planck Collaboration} {et~al.}(2016){Planck Collaboration}, {Ade},
  {Aghanim}, {Arnaud}, {Ashdown}, {Aumont}, {Baccigalupi}, {Banday},
  {Barreiro}, {Barrena}, {Bartlett}, {Bartolo}, {Battaner}, {Battye},
  {Benabed}, {Beno{\^\i}t}, {Benoit-L{\'e}vy}, {Bernard}, {Bersanelli},
  {Bielewicz}, {Bikmaev}, {B{\"o}hringer}, {Bonaldi}, {Bonavera}, {Bond},
  {Borrill}, {Bouchet}, {Bucher}, {Burenin}, {Burigana}, {Butler}, {Calabrese},
  {Cardoso}, {Carvalho}, {Catalano}, {Challinor}, {Chamballu}, {Chary},
  {Chiang}, {Chon}, {Christensen}, {Clements}, {Colombi}, {Colombo}, {Combet},
  {Comis}, {Couchot}, {Coulais}, {Crill}, {Curto}, {Cuttaia}, {Dahle},
  {Danese}, {Davies}, {Davis}, {de Bernardis}, {de Rosa}, {de Zotti},
  {Delabrouille}, {D{\'e}sert}, {Dickinson}, {Diego}, {Dolag}, {Dole},
  {Donzelli}, {Dor{\'e}}, {Douspis}, {Ducout}, {Dupac}, {Efstathiou},
  {Eisenhardt}, {Elsner}, {En{\ss}lin}, {Eriksen}, {Falgarone}, {Fergusson},
  {Feroz}, {Ferragamo}, {Finelli}, {Forni}, {Frailis}, {Fraisse}, {Franceschi},
  {Frejsel}, {Galeotta}, {Galli}, {Ganga}, {G{\'e}nova-Santos}, {Giard},
  {Giraud-H{\'e}raud}, {Gjerl{\o}w}, {Gonz{\'a}lez-Nuevo}, {G{\'o}rski},
  {Grainge}, {Gratton}, {Gregorio}, {Gruppuso}, {Gudmundsson}, {Hansen},
  {Hanson}, {Harrison}, {Hempel}, {Henrot-Versill{\'e}},
  {Hern{\'a}ndez-Monteagudo}, {Herranz}, {Hildebrandt}, {Hivon}, {Hobson},
  {Holmes}, {Hornstrup}, {Hovest}, {Huffenberger}, {Hurier}, {Jaffe}, {Jaffe},
  {Jin}, {Jones}, {Juvela}, {Keih{\"a}nen}, {Keskitalo}, {Khamitov}, {Kisner},
  {Kneissl}, {Knoche}, {Kunz}, {Kurki-Suonio}, {Lagache}, {Lamarre}, {Lasenby},
  {Lattanzi}, {Lawrence}, {Leonardi}, {Lesgourgues}, {Levrier}, {Liguori},
  {Lilje}, {Linden-V{\o}rnle}, {L{\'o}pez-Caniego}, {Lubin},
  {Mac{\'\i}as-P{\'e}rez}, {Maggio}, {Maino}, {Mak}, {Mandolesi}, {Mangilli},
  {Martin}, {Mart{\'\i}nez-Gonz{\'a}lez}, {Masi}, {Matarrese}, {Mazzotta},
  {McGehee}, {Mei}, {Melchiorri}, {Melin}, {Mendes}, {Mennella}, {Migliaccio},
  {Mitra}, {Miville-Desch{\^e}nes}, {Moneti}, {Montier}, {Morgante},
  {Mortlock}, {Moss}, {Munshi}, {Murphy}, {Naselsky}, {Nastasi}, {Nati},
  {Natoli}, {Netterfield}, {N{\o}rgaard-Nielsen}, {Noviello}, {Novikov},
  {Novikov}, {Olamaie}, {Oxborrow}, {Paci}, {Pagano}, {Pajot}, {Paoletti},
  {Pasian}, {Patanchon}, {Pearson}, {Perdereau}, {Perotto}, {Perrott},
  {Perrotta}, {Pettorino}, {Piacentini}, {Piat}, {Pierpaoli}, {Pietrobon},
  {Plaszczynski}, {Pointecouteau}, {Polenta}, {Pratt}, {Pr{\'e}zeau}, {Prunet},
  {Puget}, {Rachen}, {Reach}, {Rebolo}, {Reinecke}, {Remazeilles}, {Renault},
  {Renzi}, {Ristorcelli}, {Rocha}, {Rosset}, {Rossetti}, {Roudier}, {Rozo},
  {Rubi{\~n}o-Mart{\'\i}n}, {Rumsey}, {Rusholme}, {Rykoff}, {Sandri}, {Santos},
  {Saunders}, {Savelainen}, {Savini}, {Schammel}, {Scott}, {Seiffert},
  {Shellard}, {Shimwell}, {Spencer}, {Stanford}, {Stern}, {Stolyarov},
  {Stompor}, {Streblyanska}, {Sudiwala}, {Sunyaev}, {Sutton}, {Suur-Uski},
  {Sygnet}, {Tauber}, {Terenzi}, {Toffolatti}, {Tomasi}, {Tramonte},
  {Tristram}, {Tucci}, {Tuovinen}, {Umana}, {Valenziano}, {Valiviita}, {Van
  Tent}, {Vielva}, {Villa}, {Wade}, {Wandelt}, {Wehus}, {White}, {Wright},
  {Yvon}, {Zacchei}, \& {Zonca}}]{2016A&A...594A..27P}
{Planck Collaboration}, {Ade}, P.~A.~R., {Aghanim}, N., {et~al.} 2016, \aap,
  594, A27

\bibitem[{{Poole} {et~al.}(2006){Poole}, {Fardal}, {Babul}, {McCarthy},
  {Quinn}, \& {Wadsley}}]{2006MNRAS.373..881P}
{Poole}, G.~B., {Fardal}, M.~A., {Babul}, A., {et~al.} 2006, \mnras, 373, 881

\bibitem[{{Pratt} {et~al.}(2009){Pratt}, {Croston}, {Arnaud}, \&
  {B{\"o}hringer}}]{2009A&A...498..361P}
{Pratt}, G.~W., {Croston}, J.~H., {Arnaud}, M., \& {B{\"o}hringer}, H. 2009,
  \aap, 498, 361

\bibitem[{{Reiprich} {et~al.}(2013){Reiprich}, {Basu}, {Ettori}, {Israel},
  {Lovisari}, {Molendi}, {Pointecouteau}, \&
  {Roncarelli}}]{2013SSRv..177..195R}
{Reiprich}, T.~H., {Basu}, K., {Ettori}, S., {et~al.} 2013, \ssr, 177, 195

\bibitem[{{Rengelink} {et~al.}(1997){Rengelink}, {Tang}, {de Bruyn}, {Miley},
  {Bremer}, {Roettgering}, \& {Bremer}}]{1997A&AS..124..259R}
{Rengelink}, R.~B., {Tang}, Y., {de Bruyn}, A.~G., {et~al.} 1997, \aaps, 124,
  259

\bibitem[{{Rudnick} {et~al.}(2006){Rudnick}, {Delain}, \&
  {Lemmerman}}]{2006AN....327..549R}
{Rudnick}, L., {Delain}, K.~M., \& {Lemmerman}, J.~A. 2006, Astronomische
  Nachrichten, 327, 549

\bibitem[{{Santos} {et~al.}(2008){Santos}, {Rosati}, {Tozzi}, {B{\"o}hringer},
  {Ettori}, \& {Bignamini}}]{2008A&A...483...35S}
{Santos}, J.~S., {Rosati}, P., {Tozzi}, P., {et~al.} 2008, \aap, 483, 35

\bibitem[{{Shimwell} {et~al.}(2014){Shimwell}, {Brown}, {Feain}, {Feretti},
  {Gaensler}, \& {Lage}}]{2014MNRAS.440.2901S}
{Shimwell}, T.~W., {Brown}, S., {Feain}, I.~J., {et~al.} 2014, \mnras, 440,
  2901

\bibitem[{{van Weeren} {et~al.}(2016){van Weeren}, {Brunetti}, {Br{\"u}ggen},
  {Andrade-Santos}, {Ogrean}, {Williams}, {R{\"o}ttgering}, {Dawson}, {Forman},
  {de Gasperin}, {Hardcastle}, {Jones}, {Miley}, {Rafferty}, {Rudnick},
  {Sabater}, {Sarazin}, {Shimwell}, {Bonafede}, {Best}, {B{\^\i}rzan},
  {Cassano}, {Chy{\.z}y}, {Croston}, {Dijkema}, {En{\ss}lin}, {Ferrari},
  {Heald}, {Hoeft}, {Horellou}, {Jarvis}, {Kraft}, {Mevius}, {Intema},
  {Murray}, {Orr{\'u}}, {Pizzo}, {Sridhar}, {Simionescu}, {Stroe}, {van der
  Tol}, \& {White}}]{2016ApJ...818..204V}
{van Weeren}, R.~J., {Brunetti}, G., {Br{\"u}ggen}, M., {et~al.} 2016, \apj,
  818, 204

\bibitem[{{van Weeren} {et~al.}(2019){van Weeren}, {de Gasperin}, {Akamatsu},
  {Br{\"u}ggen}, {Feretti}, {Kang}, {Stroe}, \&
  {Zandanel}}]{2019SSRv..215...16V}
{van Weeren}, R.~J., {de Gasperin}, F., {Akamatsu}, H., {et~al.} 2019, \ssr,
  215, 16

\bibitem[{{Wang} \& {Markevitch}(2018)}]{2018ApJ...868...45W}
{Wang}, Q. H.~S. \& {Markevitch}, M. 2018, \apj, 868, 45

\bibitem[{{Wang} {et~al.}(2016){Wang}, {Markevitch}, \&
  {Giacintucci}}]{2016ApJ...833...99W}
{Wang}, Q. H.~S., {Markevitch}, M., \& {Giacintucci}, S. 2016, \apj, 833, 99

\bibitem[{{Wen} {et~al.}(2012){Wen}, {Han}, \& {Liu}}]{2012ApJS..199...34W}
{Wen}, Z.~L., {Han}, J.~L., \& {Liu}, F.~S. 2012, \apjs, 199, 34

\bibitem[{{Werner} {et~al.}(2009){Werner}, {Zhuravleva}, {Churazov},
  {Simionescu}, {Allen}, {Forman}, {Jones}, \& {Kaastra}}]{2009MNRAS.398...23W}
{Werner}, N., {Zhuravleva}, I., {Churazov}, E., {et~al.} 2009, \mnras, 398, 23

\bibitem[{{Werner} {et~al.}(2016){Werner}, {ZuHone}, {Zhuravleva}, {Ichinohe},
  {Simionescu}, {Allen}, {Markevitch}, {Fabian}, {Keshet}, {Roediger},
  {Ruszkowski}, \& {Sanders}}]{2016MNRAS.455..846W}
{Werner}, N., {ZuHone}, J.~A., {Zhuravleva}, I., {et~al.} 2016, \mnras, 455,
  846

\bibitem[{{Willingale} {et~al.}(2013){Willingale}, {Starling}, {Beardmore},
  {Tanvir}, \& {O'Brien}}]{2013MNRAS.431..394W}
{Willingale}, R., {Starling}, R.~L.~C., {Beardmore}, A.~P., {Tanvir}, N.~R., \&
  {O'Brien}, P.~T. 2013, \mnras, 431, 394

\bibitem[{{Yu} {et~al.}(2011){Yu}, {Tozzi}, {Borgani}, {Rosati}, \&
  {Zhu}}]{2011A&A...529A..65Y}
{Yu}, H., {Tozzi}, P., {Borgani}, S., {Rosati}, P., \& {Zhu}, Z.~H. 2011, \aap,
  529, A65

\bibitem[{{Zhang} {et~al.}(2019){Zhang}, {Churazov}, {Forman}, \&
  {Lyskova}}]{2019MNRAS.488.5259Z}
{Zhang}, C., {Churazov}, E., {Forman}, W.~R., \& {Lyskova}, N. 2019, \mnras,
  488, 5259

\bibitem[{{ZuHone} {et~al.}(2011){ZuHone}, {Markevitch}, \&
  {Lee}}]{2011ApJ...743...16Z}
{ZuHone}, J.~A., {Markevitch}, M., \& {Lee}, D. 2011, \apj, 743, 16

\end{thebibliography}
\bibliographystyle{aa}

\end{document}